\documentclass[onecolumn,secnumarabic,a4paper,nobibnotes,aps,prd,showpacs,showkeys,notitlepage]{revtex4-1}
\usepackage[T1]{fontenc}
\usepackage{amsmath}
\usepackage{amssymb}
\usepackage[dvipdfm]{epsfig}
\usepackage{graphicx}
\usepackage{natbib}
\usepackage[margin=4cm]{geometry}

\begin{document}

\title{Binary relations between magnitudes of different dimensions 
used in material science optimization problems. 
Pseudo-state equation of Soft Magnetic Composities}  

\author{Krzysztof Z. Sokalski}

\affiliation{Institute of Computer Science, 
Cz\c{e}stochowa University of Technology, 
Al. Armii Krajowej 17, 42-200 Cz\c{e}stochowa, Poland}

\author{Bartosz Jankowski}

\author{Barbara \'{S}lusarek}

\affiliation{Tele and Radio Research Institute, Ratuszowa 11 Street, 03-450 Warszawa, Poland} 




\begin{abstract}
New algoritm for optimizing technological parameters 
of soft magnetic compozites has been derived on the base
 of topological structure of the power loss characteristics.  
In optimization processes of magnitudes obeying scaling 
it happen binary relations of magnitudes having different dimensions. 
From mathematical point of view in general case such a procedure 
is not permissible. However, in a case of the system obeying 
the scaling law it is so.  It has been shown that in such systems 
binary relations of magnitudes of different dimensions is correct 
and has mathematical meaning which is important for practical use 
of scaling in optimization processes.  Derived here structure 
of the set of all power loss characteristics in soft magnetic 
composite enables us to derive a formal pseudo-state equation of SMC. 
This equation  constitutes a realation of the hardening temperature, 
the compaction pressure and a parameter characterizing the power 
loss characteristic. Finally, the pseudo-state equation improves 
the algoritm for designing the best values of technological parameters. 
\end{abstract}

\maketitle

\section{Introduction}\label{I}
Recently novel concept of technological parameters' optimization 
has been applied in Soft Magnetic Composities (SMC) by \'{S}lusarek et al., 
\citep{bib:slus}. This concept bases on assumption that SMC 
is a self-similar system where function of loss of power obeys 
the scaling law \citep{bib:Sokal}-\citep{bib:Sokal4}. 
The scaling is very useful tool due of the three reasons: 
\begin{itemize}
\item it reduces number of independent variables $f$ and $B_{m}$ 
to the effective one $f/B_{m}^{\alpha}$, 
\item and determines general form of loss of power characteristic 
in a form of  homogenous function in general sense (h.f.g.s.),
\item as well as enables us to use binary relations between magnitudes 
of different dimensions. 
\end{itemize}
Reduction of independent variables bases on definition of h.f.g.s. 
$F(f,B_{m})$ is h.f.g.s. if:
\begin{equation}
\label{higs}
\exists\{a,b,c\} \in \mathbf{R^{3}}: 
\forall \lambda \in \mathbf{R_{+}} \hspace{2mm} 
F(\lambda^{a}f,\lambda^{b}B_{m})=\lambda^{c}\,F(f,B_{m}).
\end{equation}
According to the assumtion concerning $\lambda$ we are free to substitute any positive real number, for instance $\lambda=B_{m}^{-\frac{1}{b}}$, then we get:
\begin{equation}
\label{general}
\frac{F(f,B_{m})}{B_{m}^{\beta}}=F\left(\frac{f}{B_{m}^{\alpha}},1\right)
\end{equation}
where $f$ and $B_{m}$ are frequency and pik of magnetic inductance, respectively. $F(\cdot,1)$ is an arbitrary function, $\alpha=\frac{a}{b}$ and $\beta=\frac{c}{b}$ are scaling exponents. 

Choice for the $F(\cdot,1)$ depends on the power loss characteristics of investigated materials. In \citep{bib:slus} we have modified the Bertotti decomposition rule \citep{bib:GB}-\citep{bib:GB2} which led to the following form for $P_{tot}(\cdot)=F(\cdot,1)$: 
\begin{widetext}
\begin{equation} 
\label {eq8} 
\frac{P_{tot}(f,B_{m})}{B_m^{\beta}}= \Gamma_{1}\,\frac{f}{B_m^{\alpha}}+ \Gamma_{2}\,\left(\frac{f}{B_m^{\alpha}}\right)^2 +
\Gamma_{3}\,\left(\frac{f}{B_m^{\alpha}}\right)^3 +\Gamma_{4}\,\left(\frac{f}{B_m^{\alpha}}\right)^4, 
\end{equation}
\end{widetext}
where $\Gamma_{n}$  have been estimated for different values of the technological parameters \citep{bib:slus} (pressure and temperature).
For purpose of this paper we take into account only one family of power loss characteristics which is presented in Fig. \ref{Fig.3} and Fig. \ref{Fig.4}. The corresponding estimated values of the model parameters are presented in TABLE \ref{Table:Table1}. For all other details concerning SMC material and measurement data we refer to \citep{bib:slus}. Now we are ready to formulate the goals of this paper.\\
Main goal is to minimize the power loss in SMC by optimal using model density of power loss (\ref{eq8}) and corresponding  
From the first row of TABLE \ref{Table:Table1} we can see that dimensions of the $\Gamma_{i}$ coefficients depend on the values of the $\alpha$ and $\beta$ exponents. Therefore, the power loss characteristics presented in Fig. \ref{Fig.3} and Fig. \ref{Fig.4} are  different dimensions. So,  we have to answer to the following quastion: are we able to relate them in the optimisation process which has been described in \citep{bib:slus}? In this paper we will prove that if the considered characteristics obey the scaling, then the binary relation between them is invariant with respect to this transformation and comparison of two magnitudes of different dimensions has mathematical meaning. \\
Reach measurement data of power losses in Somaloy 500 have been transformed into parameters of (\ref{eq8}) v.s. hardening temperature and compaction pressure Table \ref{Table:Table1} in \citep{bib:slus}. Informations contained in this Table  enable us to infer about topological structure of set of the power loss characteistics and finally to construct pseudo-state equation for SMC, and derive supplementary algoritm for the best values of technological parameters to that one published in \citep{bib:slus}.   
\section{Scaling of binary relations}\label{II}
Let the power loss characteristic has the form determined by the scaling (\ref{general}). It is important to remain that $\alpha$ and $\beta$ are defined by initial exponents $a,b$ and $c$ (see after formula (\ref{general})): 
\begin{equation}
\label{wykladniki}
\alpha=\frac{a}{b},\hspace{2mm}\beta=\frac{c}{b}.
\end{equation}
Let us concentrate our attention at the point on the $\frac{f}{B_m^{\alpha}}$ axis of Fig. \ref{Fig.3} or Fig. \ref{Fig.4}:
\begin{equation}
\frac{f}{B_m^{\alpha}}=\frac{f_{1}}{B_{m\,1}^{\alpha_{1}}}=\frac{f_{2}}{B_{m\,2}^{\alpha_{2}}}=\cdots=\frac{f_{4}}{B_{m\,4}^{\alpha_{4}}}
\end{equation}
Let us take into account the two characteristics and let us assume that
\begin{equation}
\label{ineq1}
\frac{P_{tot\,1}}{B_{m\,1}^{\beta_{1}}}>\frac{P_{tot\,2}}{B_{m\,2}^{\beta_{2}}}.
\end{equation}
Therefore, the considered binary relation is the strong inequality and corresponds to natural order presented in Fig. \ref{Fig.3} and Fig. \ref{Fig.4}. The most important question of this research is whether (\ref{ineq1}) is invariant with respect to scaling:
\begin{equation}
\label{ineq2}
\frac{P_{tot\,1}'}{B_{m\,1}'^{\beta_{1}}}>\frac{P_{tot\,2}'}{B_{m\,2}'^{\beta_{2}}}.
\end{equation}
 Let $\lambda>0$ be an arbitrary positive real number. Then, the scaling of (\ref{ineq2}) goes according to the following algoritm:
\begin{itemize}
\item Let us perorm the scaling with respect to $\lambda$ of all independent magnitudes and the dependent one :
\begin{eqnarray}
f_{i}'=\lambda^{a_{i}}f_{i},\nonumber\\
B_{m\,i}'=\lambda^{b_{i}}B_{m\,i},\label{skal1}\\
P_{tot\,i}'=\lambda^{c_{i}}P_{tot\,i},\nonumber
\end{eqnarray}
where $i=1,2\cdots 4$ labels the considered characteristics.
\item Substituting apropriate relations of (\ref{skal1}) to (\ref{ineq2}) we derive:
\begin{equation}
\label{scal2}
\frac{P_{tot\,1}}{B_{m\,1}^{\beta_{1}}}\lambda^{c_{1}-b_{1}\beta_{1}}>\frac{P_{tot\,2}}{B_{m\,2}^{\beta_{2}}}\lambda^{c_{2}-b_{2}\beta_{2}}.
\end{equation}
\item Collecting all powers of $\lambda$ on the left-hand side of  (\ref{scal2}) and  taking into account (\ref{wykladniki}) we derive the resulting power to be zero and:
\begin{equation}
\lambda^{c_{1}-c_{2}-b_{1}\beta_{1}+b_{2}\beta_{2}}=1.
\end{equation}
\end{itemize}
Therefore (\ref{ineq1}) is invariant with respect to scaling. This binary relation has mathematical meaning and constitutes the total order in the set of characteristics.
\section{Binary Equivalence Relation}\label{III}
The result derived in Section \ref{II} can be supplement with the following binary equivalence relation. Let 
\begin{equation}
\label{point}
X_{i,j}=\left(\frac{f_{i,j}}{B_{m\,i,j}^{\alpha_{j}}},\frac{P_{tot\,i,j}}{B_{m\,i,j}^{\beta_{j}}}\right)
\end{equation}
 be $i$-th pont of the $j$-th characteristic. Two  points are related if they belong to the same characteristic:
\begin{equation}
\label{relat1}
X_{i,j}\,\mathcal{R}\,X_{k,j}.
\end{equation}
{\em Theorem:} $\mathcal{R}$ is equivalence relation. (The proof is trivial and can be done  by checking out that the considered relation is: reflexive, symmetric and  transitive.) Therefore, $\mathcal{R}$ constitutes division of the positive-positive quarter of plane spanned by (\ref{point}). The characteristics do not intersect each other except in the origin point which is excluded from the space. The result of this section implies that the power loss characteristics (\ref{general}) and (\ref{eq8}) are invariant with respect to scaling. 


\begin{table}
\scriptsize
\caption {Somaloy 500. Values of scaling exponents and coefficients of (\ref{eq8}) v.s. compaction pressure and hardening temperatures, a selection from \citep{bib:slus}}
\label{Table:Table1}
    \begin{tabular}{|c|c|c|c|c|c|c|c|}
    \hline
Temperature & Pressure & $\alpha$ & $\beta$ & $\Gamma_{1}$ & $\Gamma_{2}$ & $\Gamma_{3}$ & $\Gamma_{4}$ \\
	\hline
$[{}^{o}{C}]$ & $[MPa]$ & $[-]$ & $[-]$ & $[\frac{m^{2}}{s^{2}}T^{\alpha-\beta}]$ & $[\frac{m^{2}}{s}T^{2\alpha-\beta}]$ & $[m^{2}T^{3\alpha-\beta}]$ & $[{m^{2}}{s}\,T^{4\alpha-\beta}]$\\
    \hline
		500	&500	&-1.312	&-0.011	&0.171	&$3.606 \cdot 10^{-5}$	&$1.953 \cdot 10^{-8}$	&$-2.255 \cdot 10^{-12}$\\
		500	&600	&-1.383	&-0.125	&0.153	&$3.328 \cdot10^{-5}$	&$9.254 \cdot 10^{-9}$	&$-1.177 \cdot 10^{-12}$\\
		500	&700	&-1.735	&-0.517	&0.156	&$2.393 \cdot 10^{-5}$	&$2.309 \cdot 10^{-9}$	&${-8}.075 \cdot 10^{-14}$\\
		500	&900	&-1.395	&-0.082	&0.101	&$6.065 \cdot 10^{-5}$	&${-8}.031 \cdot 10^{-9}$	&$7.877 \cdot 10^{-13}$\\
		400  &800 &-1.473 &-0.28  &0.183 &$1.347\cdot 10^{-5}$   &$3.689 \cdot 10^{-9}$        &$1.185 \cdot 10^{-13}$\\
		450  &800 &-1.596 &-0.123&0.145 &$2.482\cdot 10^{-5}$   &$-1.218 \cdot 10^{-9}$	  &$6.120 \cdot 10^{-14}$\\
		550  &800 &-2.034 &-1.326&0.106 &$1.407\cdot 10^{-4}$  &$-1.066\cdot 10^{-8} $    &$4.541\cdot 10^{-13}$\\
		600  &800&-1.608 & -0.232&1.220 &$8.941\cdot 10^{-4}$  &$-5.302\cdot 10^{-8}$     &$1.664\cdot 10^{-11}$\\
			\hline
    \end{tabular}
\end{table}
\begin{figure}
\begin{center}
\includegraphics[ width=10cm]{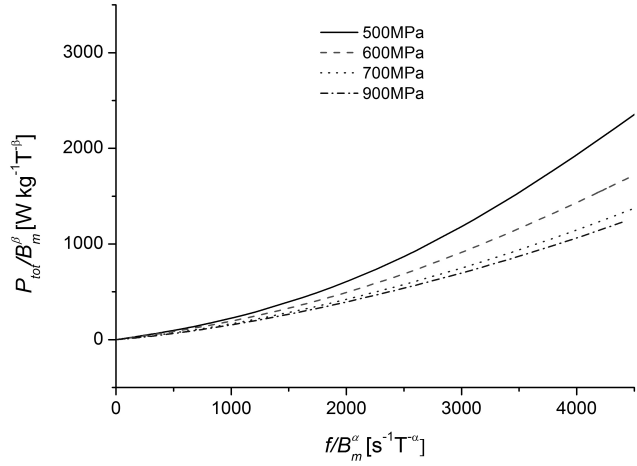}
\caption{Selection of the power loss characteristics  
$P_{tot}/B_{m}^{\beta}$ vs. $f/B_{m}^{\alpha}$ calculated according 
to (\ref{eq8}) and Table \ref{Table:Table1} for Somaloy 500 \citep{bib:slus}. 
Corresonding hardening temperature was $500^{o}C$.}
\label{Fig.3}
\end{center}
\end{figure} 
\begin{figure}
\begin{center}
\includegraphics[ width=10cm]{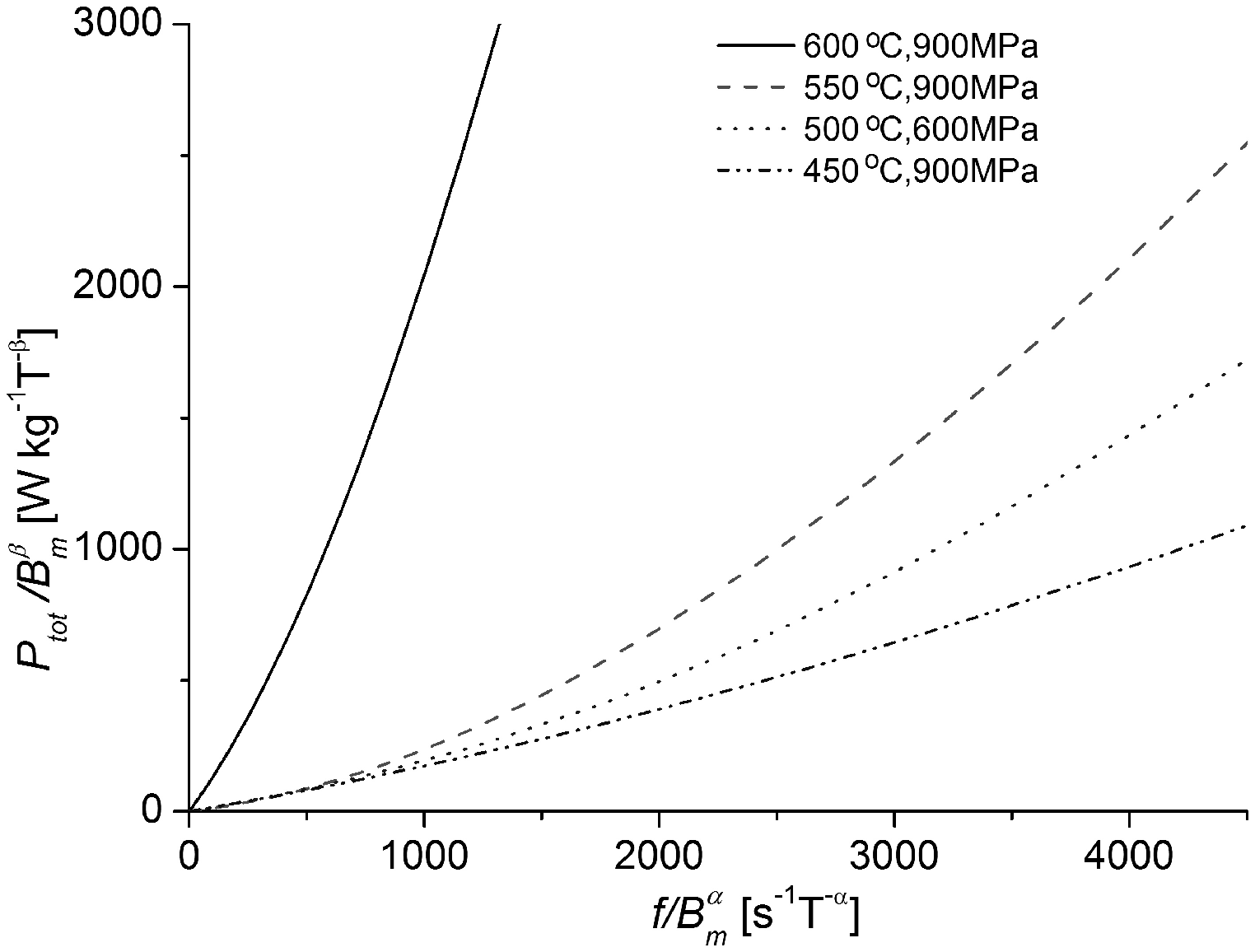}
\caption{A selection of the power loss characteristics  
$P_{tot}/B_{m}^{\beta}$ vs. $f/B_{m}^{\alpha}$ calculated according to (\ref{eq8}) and Table \ref{Table:Table1} for Somaloy 500 \citep{bib:slus}.
}
\label{Fig.4}
\end{center}
\end{figure} 
 Structure of derived here the set of all characteristics of which some examples are presented in Fig. \ref{Fig.3} and Fig. \ref{Fig.4} enables us to derive a formal pseudo-state equation of SMC. This equation  constitutes a realation of the hardening temperature, the compaction pressure and a parameter characterizing the power loss characteristic corresponding to the values of these technological parameters. Finally, the pseudo-state equation will improve the algoritm for designing the best values of technological parameters. 
\section{Pseudo-state equation of SMC}
\begin{figure}
\begin{center}
\includegraphics[ width=10cm]{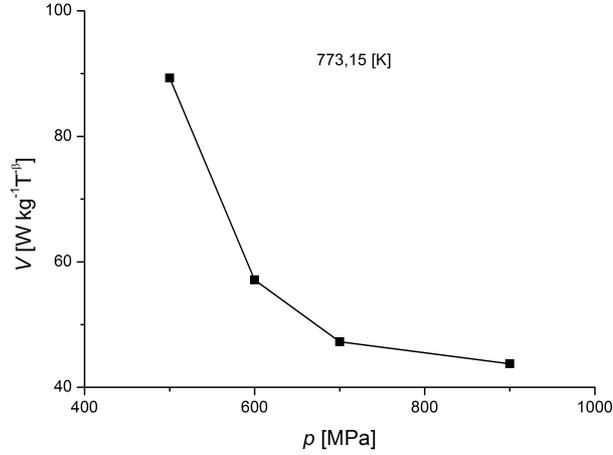}
\caption{Pseudo-Isotherm $T=500^{o}C$ of the Low-losses phase, according to data of Table \ref{Table:Table2}}. 
\label{Fig.5}
\end{center}
\end{figure} 
Let $\tilde{C}$ be set of all possible power loss characteristics in considered SMC. Each characteristic is smooth curve in $\left[f/B_{m}^{\alpha},P_{tot}/B_{m}^{\beta}\right]$ plane which corresponds to a point in $\left[ {T}, {p}\right]$ plane. 
In order to derive the pseudo-state equation we transform each power loss characteristic into a number $V$ corresponding to $\left( {T}, {p}\right)$ point. By this way we obtain a function of two variables:
\begin{equation}
\label{eq01}
\left( {T}, {p}\right)\longrightarrow V.
\end{equation}
This function must satisfy the following condition.   
Let us concentrate our attantion at the two following points: 
\begin{eqnarray}
\frac{f_{1}}{B_{m\,1}^{\alpha_{1}}}=\frac{f}{B_{m}^{\alpha}}\nonumber\\
\frac{f_{2}}{B_{m\,2}^{\alpha_{2}}}=\frac{f}{B_{m}^{\alpha}}\label{ineq01}.
\end{eqnarray}
Let us consider the two characteristics $P_{tot\,1}/ {B_{m\,1}^{\beta_{1}}} $ and $P_{tot\,2}/ {B_{m\,2}^{\beta_{2}}} $ of the two samples composed under $ {T_{1}}, {p_{1}}$ and $ {T_{2}}, {p_{2}}$ values of temperature and pressure, respectively. While, the other technological parameters powder compositions and  volume fraction are constant. Let us assume that for (\ref{ineq01}) the following relation holds:
\begin{equation}
\label{ineq11}
\frac{P_{tot\,1}}{B_{m\,1}^{\beta_{1}}}>\frac{P_{tot\,2}}{B_{m\,2}^{\beta_{2}}}.
\end{equation}
It results from the derived structure of $\tilde{C}$ that (\ref{ineq11}) holds for each value of (\ref{ineq01}). Therefore we have to assume the following condition of sought $V( {T}, {p})$:  If the relation (\ref{ineq11}) holds for $ {T_{1}}, {p_{1}}, {T_{2}}, {p_{2}}$  then the following relation has to be satisfied for $V( {T}, {p})$:
\begin{equation}
\label{ineq22}
V( {T_{1}}, {p_{1}})>V( {T_{2}}, {p_{2}}).
\end{equation}
Moreover, $V( {T}, {p})$ has to indicate place of coresponding characteristic in the ordered $ \tilde{C}$. The simplest choice satisfying these requirements is the following average:
\begin{equation}
\label{V1}
V( {T}, {p})=\frac{1}{\phi_{max}-\phi_{min}}\int_{\phi_{min}}^{\phi_{max}}\frac{P_{tot}(\frac{f}{B_{m}^{\alpha}})}{B_{m}^{\beta}}\,d(\frac{f}{B_{m}^{\alpha}}),
\end{equation}
where the integration domain is common for the all characteristics. We have selected the common domain of Fig. \ref{Fig.3} and Fig. \ref{Fig.4}: $\phi_{min}=0,\hspace{2mm} \phi_{max}=4000 [\frac{1}{s\,T^{\alpha}}]$.  Using (\ref{eq8}) we transform (\ref{V1}) to the working formula for the measure $V$:
\begin{equation}
\label{V2}
V( {T}, {p})=\frac{1}{\phi_{min}-\phi_{max}}\int_{\phi_{min}}^{\phi_{max}}\,x\,(\Gamma_{1}+x\,(\Gamma_{2}+x\,(\Gamma_{3}+x\,\Gamma_{4})))\,dx,
\end{equation}
where, $x=\frac{f}{B_{m}^{\alpha}}$.  $\alpha,\Gamma_{i}$ are  coefficients dependent on $ {T}$ and $ {p}$, see TABLE \ref{Table:Table1}. The values of  $V( {T}, {p})$ are tabulated in TABLE \ref{Table:Table2}.
\begin{table}
\scriptsize
\caption {$V$ measure v.s. hardening temperature and compaction pressure.}
\label{Table:Table2}
    \begin{tabular}{|c|c|c|}
    \hline
Temperature & Pressure & V \\
	\hline
$[K]$ & $[MPa]$ & $\left[\frac{W}{kg\,T^{\beta}}\right]$ \\
    \hline
723,15&	800&	40,6\\
773,15&	900&	43,75\\
773,15&	700&	47,25\\
673,15&	800&	50,3\\
773,15&	600&	57,12\\
823,15&	800&	81,5\\
773,15&	500&	89,28\\
\hline
742,15&	764&	492,3\\
753,15&	780&	509,2\\
804,15&	764&	528,5\\
711,15&	764&	547,0\\
873,15&	800&	720,0\\
\hline
    \end{tabular}
\end{table}
TABLE \ref{Table:Table2} enables us to draw pseudo-izotherm. It is presented in Fig. \ref{Fig.5}.
However, in order to derive the complete pseudo-state equation we must create a mathematical model. On basis of Fig. \ref{Fig.5}  we start from the classical gas state-equation as an initial approximation:
\begin{equation}
\label{V3}
\frac{ {p}\,V}{k_{B} {T}}=1,
\end{equation}
where $k_{B}$ is pseudo-Boltzmann constant. In order to extent (\ref{V3}) to a realistic equation we apply again the scaling hypothesis (\ref{general}) \citep{bib:Sokal}-\citep{bib:Sokal4}:
\begin{equation}
\label{general2}
V\left(\frac{T}{T_{c}},\frac{p}{p_{c}}\right)=\left(\frac{p}{p_{c}}\right)^{\gamma}\,\cdot\,\Phi\left(\frac{\frac{T}{T_{c}}}{(\frac{p}{p_{c}})^{\delta}}\right),
\end{equation}
where $\Phi(\cdot)$ is an arbitrary function to be determined.  $\gamma$, $\delta$  and $T_{c}$, $p_{c}$ are scaling exponents and scaling parameters respectively, to be determined. For our conveniences we introduce the following variables:
\begin{equation}
\tau=\frac{T}{T_{c}},
\hspace{2mm}\pi=\frac{p}{p_{c}},
\hspace{2mm}X=\frac{\frac{T}{T_{c}}}{(\frac{p}{p_{c}})^{\delta}}
=\frac{\tau}{\pi^{\delta}}.
\end{equation}
In order to extent (\ref{V3}) to a full state-equation we apply  Pad\'e approximante by analogy to virial expantion derived by Ree and Hoover \citep{bib:Ree}:
 \begin{equation} 
\label {eq8a} 
V(\tau,\pi)= \pi^{\gamma}\frac{G_{0}+G_{1}\,X+ G_{2}\,X^2 +
G_{3}\,X^3 +G_{4}\,X^4}{1+D_{1}\,X+ D_{2}\,X^2 +
D_{3}\,X^3 +D_{4}\,X^4}, 
\end{equation}
where  $G_{0},\dots,G_{4}, D_{1},\dots,D_{4}$ are parameters of the Pad\'e approximante. All parameters
have to be determined from the data presented in TABLE \ref{Table:Table2}. 
\section{Estimation of the psudo-state equation's parameters}
At the beginning we have to notice that the data collected in TABLE \ref{Table:Table2} reveal sudden change of $V$ between two points:
$[773,15;500,0]$ and $[742,15;764,0]$. This suggests existence of a crossover between two phases: low-losses phase  and high-losses phase. We take this effect into account and we divide the data of Table \ref{Table:Table2} into two subsets corresponding to these two phases, respectively. Since the crossover consists in changing of characteristic exponents for the given universality class it is necessary to perform estimations of the   model  parameters for  each phase separately. Minimizations of $\chi^{2}$ for both phases have been performed by using MICROSOFT EXCEL 2010, where
\begin{equation}
\label{chi2}
\chi^{2}=\sum_{i=1}^{N}\left( V(\tau_{i},\pi_{i})- \pi_{i}^{\gamma}\frac{G_{0}+G_{1}\,X_{i}+ G_{2}\,X_{i}^2 +
G_{3}\,X_{i}^3 +G_{4}\,X_{i}^4}{1+D_{1}\,X_{i}+ D_{2}\,X_{i}^2 +
D_{3}\,X_{i}^3 +D_{4}\,X_{i}^4}\right)^{2}, 
\end{equation}
where $N=7$ and $N=5$ for the low-losses and high-losses phases, respectively.
  Table \ref{Table:Table3} and Table \ref{Table:Table4} present estimated values of the model parameters for the low-losses  and for high-losses phases, respectively.

\begin{table}
\scriptsize
\caption {Somaloy 500, low-losses phase.  Values of pseudo-state equation's parameters and the Pad\,{e} aproximant's coefficients of (\ref{eq8a}). }
\label{Table:Table3}
    \begin{tabular}{|c|c|c|c|c|c|c|}
    \hline
$\gamma$ & $\delta$ & $T_{c}$ & $p_{c}$ & $G_{0}$ & $G_{1}$ & $G_{2}$\\
\hline
0,1715 & 1,2812 & 21,622	 & 37,729 & 370315315 & -47752251	& 1734952\\
\hline\hline
$G_{3}$&$G_{4}$& $D_{1}$ &$D_{2}$ &$D_{3}$ &$D_{4}$ & -\\ 
\hline
-1,3764 &	-678,26 & 170,80 & 6243,8 &	386,96  &	-28,699 & - \\
\hline
  \end{tabular}
\end{table} 

\begin{table}
\scriptsize
\caption {Somaloy 500, high-losses phase.  Values of pseudo-state equation's parameters and the Pad\,{e} aproximant's coefficients of (\ref{eq8a}). }
\label{Table:Table4}
    \begin{tabular}{|c|c|c|c|c|c|c|}
    \hline
$\gamma$ & $\delta$ & $T_{c}$ & $p_{c}$ & $G_{0}$ & $G_{1}$ & $G_{2}$\\
\hline
0,1810	 & 1,5550 & 22,949 & 30,197 & 365210688 &	-47714207 &	1762773\\
\hline\hline
$G_{3}$&$G_{4}$& $D_{1}$ &$D_{2}$ &$D_{3}$ &$D_{4}$ & -\\ 
\hline
-1,3763 &	-683,38 & 170,77 & 	5739,9 &	387,81 &	-22,514 & - \\
\hline
  \end{tabular}
\end{table} 
\section{Optimization of Technological Parameters}
\begin{figure}
\begin{center}
\includegraphics[ width=10cm]{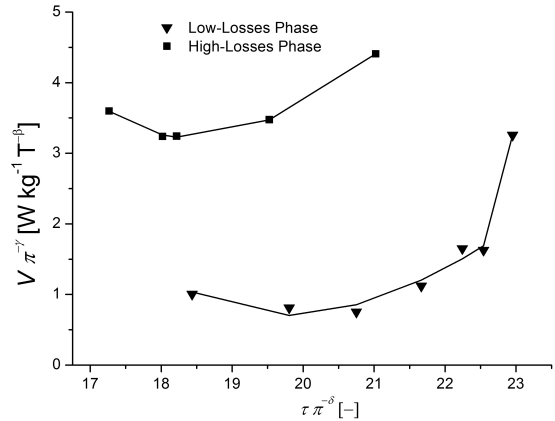}
\caption{Phase diagram for Somaloy 500.}
\label{Fig.8}
\end{center}
\end{figure}

Function $V( {T}, {p})$ serves a power loss measure versus the hardening temperature and compaction pressure. In order to explain how to optimize the technological parameters                  
with the pseudo-state equation (\ref{eq8a}) we plot the phase diagram of considered SMC Fig. \ref{Fig.8}. Note that all losses' characteristics collapsed to a one curve for the each phase. Taking into account the Low-losses phase we determine the lowest losses at $ \tau\cdot\pi^{-\delta}=19,75$. This gives the following continous subspace of the optimal points:
\begin{equation}
\label{optim}
\frac{\frac{T}{T_{c}}}{(\frac{p}{p_{c}})^{\delta}}=19,75.
\end{equation}
(\ref{optim}) represents the minimal iso-power losse curve. All points satisfying (\ref{optim}) are solutions of the optimization problem for technical parameters of SMC.

\section{Conclusions}
By introducing the binary relations we have revealed twofold. 
The power loss characteristics do not cross each other which makes 
the topology's set of this curves very useful and effectively 
that we can perform all calculations in the one-dimension space 
spanned by the scaled frequency \citep{bib:slus}-\citep{bib:Sokal4} 
or here in the case of pseudo-state equation in the scaled temperature. 
For general knowledge concerning such a topology we refer 
to the papers by Egenhofer \citep{bib:Egen1} 
and by Nedas et al. \citep{bib:Egen2}. 
However, to our knowledge this paper is the first one about 
the binary relations between magnitudes of different dimensions 
in the sense of different physical magnitudes.

The efficiency of scaling in solving problems concerning power 
losses in soft magnetic composities has already been confirmed 
in recent paper \citep{bib:slus}. However, this paper is the first 
one which presents an application of scaling in designing 
the technological parameters' values by using the pseudo-state 
euation of SMC. The obtained result is the continuous set 
of points satisfying (\ref{optim}). All solutions of this equations 
are equivalent for the optimization of the power losses. 
Therefore, the remaining degree of freedom can be used 
for optimizing magnetic properties of the considered SMC.  
Ultimately, one must say that the degree of success achieved 
when applying the scaling depends on the property of the data.
The data must obey the scaling. 

\begin{acknowledgments}
The work has been supported by National Center of Science 
within the framework of research project Grant N N507 249940.
\end{acknowledgments}

\bibliographystyle{plainnat}
 
\end{document}